\documentclass{emulateapj}

\newcommand{\beq}{\begin{equation}}
\newcommand{\beqa}{\begin{eqnarray}}
\newcommand{\eeq}{\end{equation}}
\newcommand{\eeqa}{\end{eqnarray}}

\begin{document}
\title{The Remnants of Intergalactic Supernovae}

\author{Dan Maoz\altaffilmark{1}, Eli Waxman\altaffilmark{2}, Abraham
  Loeb\altaffilmark{3,4}}

\altaffiltext{1} {School of Physics and Astronomy, 
Tel Aviv University, Tel
Aviv 69978, ISRAEL; dani@wise.tau.ac.il}

\altaffiltext{2} {Physics Department, Weizmann Institute, Rehovot,
70100, ISRAEL; waxman@wicc.weizmann.ac.il}

\altaffiltext{3} {Astronomy Department, Harvard University, 60 
Garden Street, Cambridge, MA 02138; aloeb@cfa.harvard.edu}

\altaffiltext{4} {Einstein Minerva Center, Weizmann Inst. of Science}

\begin{abstract}
Intergalactic type-Ia supernovae (SNe) have been discovered recently in
rich galaxy clusters. These SNe are likely the descendants of an
intergalactic stellar population, which has been discovered in recent years
through a variety of tracers.  We estimate the observational signatures of
the associated SN remnants (SNRs) in the unusual intracluster medium (ICM)
environment.  We find that, if type-Ia SNe still have a circumstellar medium
(CSM) at the time of explosion, then their remnants are visible in the
optical for $\sim 10^{2-3}$~years, with properties similar to young
galactic SNRs. In contrast with galactic SNRs, in which the ejecta from the
explosion interacts with the interstellar medium (ISM), intracluster SNRs
become undetectable in the optical band once their ejecta passes beyond the
CSM and enters the hot and tenuous ICM. If type-Ia SNe have a CSM, there
should be $\sim 150$ young SNRs in the nearby Virgo cluster, with an
H$\alpha$ luminosity of $\sim 10^{35}$~erg~s$^{-1}$ and an
angular size $\sim 0.1''$. We investigate the possibility
 that members of this SNR population
may have recently been detected, but incorrectly identified as
intergalactic H~II regions. Alternatively, if optical intergalactic SNRs do
not exist in Virgo, this will constitute evidence that type-Ia SNe are
devoid of a CSM, with implications for progenitor scenarios. Regardless of
the presence of a CSM, about 10 older SNRs per square degree should be
detectable in Virgo in the radio band, with fluxes of order 0.1~mJy at
1~GHz. Their angular sizes, $\sim1''$, morphologies, and lack of optical
association with distant galaxies can distinguish them from the much more
numerous background population. Their detection would provide an accurate
measurement of the intracluster SN rate. Deep pointed observations
toward the site of SN1980I, a possibly intergalactic type-Ia event in
Virgo, could test for the existence of a CSM by comparison to our
predictions for the early time development of intergalactic SNRs.
\end{abstract}

\keywords{supernova remnants --- supernovae: general --- galaxies:
  clusters: general}

\section{Introduction}
Studies of nearby clusters have revealed a population of intergalactic
stars. Already noticed by Zwicky (1951) as excess starlight
between the galaxies in the core of the Coma cluster, this diffuse stellar
emission has been confirmed and quantified in recent deep images of Coma
(Gregg \& West 1998; Trentham \& Mobasher 1998; Feldmeier et al.  2002), in
other nearby clusters (Calc{\' a}neo-Rold{\' a}n et al. 2000), and in
stacked images of redshift 
$z\sim 0.25$ clusters from the Sloan Digital Sky Survey 
(Zibetti et al. 2005).  Intergalactic red giant
stars have been detected in the Virgo Cluster (Ferguson, Tanvir, \& von
Hippel 1998; Durrell et al. 2002), and intergalactic planetary nebulae have
been found in Virgo and Fornax (Arnaboldi et al. 1996; Theuns \& Warren
1997; Mendez et al. 1997; Ciardullo et al. 1998, 2002; Feldmeier,
Ciardullo, \& Jacoby 1998).  Some 10--20\% of the stars in galaxy clusters
are in the intergalactic component.  This population is believed to have
been stripped off the cluster galaxies through tidal disruption by other
galaxies and by the cluster potential as a whole (Dubinski, Mihos, \&
Hernquist 1996; Moore et al. 1996; Korchagin, Tsuchiya, \& Miyama 2001).

In the course of a survey for supernovae (SNe) in rich
galaxy clusters at redshifts $0.08<z<0.2$, Gal-Yam et al. (2003) 
recently discovered
two type-Ia SNe at the redshifts of their respective clusters, but
spatially and and kinematically distinct from any galaxy in the
cluster. The two events, constituting 2/7 of the cluster SNe found in the
survey, had no detectable host galaxy, even in deep images taken by the
Keck 10~m telescope.  Gal-Yam et al. argued, based on the galaxy luminosity
function of clusters, that dwarf-galaxies below the detection limit, and
contributing only $\sim 10^{-3}$ of the cluster stellar luminosity,
could not plausibly be the hosts of the two SNe. Accounting for the
relative detection efficiencies of events within and outside galaxies,
Gal-Yam et al. estimated that $21^{+18}_{-14}$ percent of the SN~Ia parent
stellar population in clusters is intergalactic. This fraction is
consistent with the intergalactic stellar fraction found by other
tracers. 

Quantifying the properties of the intergalactic stellar population
via its different tracers is important for understanding galaxy
interactions and evolution in dense environments. SNe are particularly
useful because, as opposed to other tracers, they can be seen out to
clusters at large lookback times, and can thus reveal the history of the
galaxy evolution process.

Supernovae are brief optical events.
Supernova remnants (SNRs), however, exist for thousands of
years and are detectable over a wide range of wavelengths. The
intergalactic SN population in clusters could hence potentially be detected
and characterized by means of the SNRs it leaves behind.  In this paper, we
predict the observational signatures of SNRs in the unusual intracluster
medium (ICM) environment.  We show that the SN ejecta is visible via its
interaction with the surrounding circumstellar medium (CSM), if the latter
exists, over a timescale of $\sim 10^2$---$10^3$ years. In galactic
environments, after traversing the CSM, the ejecta interacts with a
galaxy's interstellar medium (ISM) and the SNR remains visible for
$10^4$~yr, or more. However, in cluster environments, once the ejecta
traverses the CSM and enters into the hot and tenuous intracluster medium
(ICM), the SNR emission fades. We will thus argue that the emission from
the remnants of intergalactic type-Ia SNe is detectable only if they are
surrounded by a dense CSM (implying progenitors with a giant-star companion
that had a high mass loss rate; see review by Branch et al. 1995), and if
the remnants are not much older than a thousand years.  A CSM has been
detected in only one type-Ia SN, 2002ic, where hydrogen lines have appeared
in the late-time spectra of the explosion (Hamuy et al. 2003). However, it
is unclear if this single event is representative of the physical
conditions of most SNe-Ia which (by definition) have no signs of hydrogen
in their spectra. Thus, detection of intergalactic SN-Ia remnants could
demonstrate the existence of a CSM around type-Ia SN progenitors.
%In terms
%of numbers and properties, we will argue that intracluster SNRs may have
%already been discovered in the optical, but incorrectly identified as
%intergalactic H~II regions, powered by massive young stars.

In \S2 we use some simple physical arguments to derive the main observed
features of SNRs and their dependence on progenitor properties and on the
peculiar ICM environment.  In \S3, we review the characteristics of the
unresolved intergalactic emission-line objects recently discovered in
nearby clusters and groups, and discuss whether some or all of these
objects could be intergalactic SNRs, rather than H~II regions.  In \S4, we
examine whether radio observations could detect the intergalactic SNR
population. We summarize our results in \S5.

\section{The appearance of a SNR in the ICM}

\subsection{Wind dynamics}

We will construct a simple physical model for SNRs that reproduces
their main observed features, and will allow prediction of the
properties of intergalactic SNRs. For a detailed treatment of the
physics of SNRs, see, e.g., McCray \& Wang(1996), or Truelove \& McKee
(1999).

We consider a type-Ia SN explosion due to a white dwarf that
exceeds the Chandrasekhar mass through accretion of wind material from a
giant star companion. The giant is assumed to have a mass loss rate of
${\dot M}=10^{-6}{\dot M}_{-6} M_\odot~{\rm yr}^{-1}$, which creates a CSM
with which the SN ejecta eventually collides. Only later does the ejecta
reach the external medium, be it the ISM in the case of normal galactic
SNe, or the ICM for the intracluster SNe under discussion.

First, we consider the case where the binary star system is at rest
relative to an ambient ICM with a pressure $p_{\rm ICM}$. In this case, the
entire mass lost by the giant star accumulates in a shell around the
binary. This case applies when the speed of the binary system relative to
the ICM, $v_*$, is much smaller than the giant star's wind speed,
$v_w=10^{1.5} v_{w,1.5}~{\rm km~s^{-1}}$.

The outgoing wind acts as a piston and generates a forward shock in the
external medium (e.g., Parker 1963). 
Since the wind velocity is typically much smaller than the
sound speed of the surrounding ICM, the forward shock is weak. The
interaction with the ICM sends a reverse shock back into the cold wind
material. The shell of material trapped between the forward and reverse
shocks includes wind material separated from shocked ICM material by a
contact discontinuity. If this shell is moving forward with a velocity $v$,
then the reverse shock moves backward relative to the unperturbed wind at a
speed of,
\begin{equation}
v_s={(\gamma +1)\over 2}\left(v_w-v\right),
\label{eq:first}
\end{equation}
where $\gamma=5/3$ is the adiabatic index of a monoatomic gas.  The pressure
in this shell satisfies
\begin{equation}
\label{eq:second}
p_s={2\over (\gamma + 1)}\rho_w v_s^2 \approx p_{\rm ICM},
\end{equation}
implying a reverse shock speed relative to the ICM:
\begin{equation}
v_{\rm rs}=(v_w - v_s)=v_w\left(1-\sqrt{(\gamma+1)p_{\rm ICM}\over 2
\rho_wv_w^2}\right) .
\label{eq:v_rs}
\end{equation}
Here, $\rho_w \propto R^{-\eta}$ is the density of the wind material,
with $\eta=2$ if the star had a constant mass loss rate.
For such a density profile, 
the reverse shock turns into a standing shock in the observer's frame
at the radius $R=R_{\rm rs}$ where $v_{\rm rs}=0$, namely $v_w=v_s$. 
Setting $v_{\rm rs}=0$ in equation (\ref{eq:v_rs}), we have
\begin{eqnarray}
\nonumber
R_{\rm rs}&=&\left({{\dot M}v_w\over 2\pi(\gamma+1)p_{\rm ICM}}\right)^{1/2}
\\&&=0.6\times 10^{18}~{\dot M}_{-6}^{1/2}
v_{w,1.5}^{1/2}n_{-3}^{-1/2}T_{1}^{-1/2}~{\rm cm} ,
\end{eqnarray}
where $\rho_w\equiv n_w m_p={\dot M}/4\pi R^2 v_w$ is the mass density in
the wind (with $m_p$ being the mean particle mass), $T_{\rm ICM}=
10T_1~{\rm keV}$ is the ICM temperature, and $n_{\rm
ICM}=10^{-3}n_{-3}~{\rm cm^{-3}}$ is the electron density in the ICM (with
$p_{\rm ICM}\approx 2n_{\rm ICM}kT_{\rm ICM}$). At this point the wind
velocity is reduced to
\begin{equation}
v=\left({\gamma-1\over \gamma + 1}\right)v_w={1\over 4} v_w.
\end{equation}

The density profile beyond $R_{\rm rs}$ can be derived by the following argument.
Once a wind fluid element crosses the reverse shock, its entropy is conserved
and its pressure is approximately time independent and equal to the ICM pressure.
This implies that the density of the fluid is independent of time and radius at 
$R>R_{\rm rs}$, $\rho_w(R>R_{\rm rs})=4\dot{M}/4\pi R_{\rm rs}^2v_w=2(\gamma+1)p_{\rm ICM}/v_w^2$,
\begin{equation}\label{eq:rhow}
    \rho_w(R>R_{\rm rs})=9.2 m_p\frac{n_{-3}T_{1}}{v_{w,1.5}^2}\,{\rm cm^{-3}}.
\end{equation}
The radius independent density implies a deceleration of the wind velocity beyond $R_{\rm rs}$, 
$v_w(R>R_{\rm rs})\propto R^{-2}$. The outer radius of the wind is determined by mass conservation,
$4\pi R_{\rm max}^3\rho_w(R>R_{\rm rs})/3=\dot{M}\tau$,
\begin{equation}\label{eq:R_max}
    R_{\rm max}=\left(\frac{3}{4}R_{\rm rs}^2v_w\tau\right)^{1/3}
    =3.0\times10^{18}\left(\frac{{\dot M}_{-6}v_{w,1.5}^2\tau_6}{n_{-3}T_{1}}\right)^{1/3}{\rm cm}.
\end{equation}
Here, $\tau= 10^6\tau_6~{\rm yr}$ is the duration of the period over which
the wind is active before the SN explodes. For typical parameters,
most of the wind material is trapped between this radius and the 
stationary radius of the reverse shock, $R_{\rm rs}$.

In the ICM of massive clusters, an intergalactic stellar binary
will typically move at a high speed of $v_{\rm star} \sim 10^3~{\rm
km~s^{-1}}$ relative to the ICM. Under these circumstances, the wind
remnant will be stripped by the ram pressure of the ICM in its rest
frame. We define an effective "ram-pressure" speed, $v_\star=\sqrt{v_{\rm
star}^2+v_{\rm th}^2}=10^3v_{*,3}~{\rm km~s^{-1}}$, where $v_{\rm
th}^2=p_{\rm ICM}/\rho_{\rm ICM}$ is the thermal speed of the ICM gas.  The
wind remnant (or equivalently, the CSM) will have the same structure as
before, except that the ICM temperature, $T$, is increased by a
factor $\sim (v_\star/v_{\rm th})^2$ to $T_*=(1+ v_{\rm star}^2/v_{\rm
th}^2)T$. This leads to modification of the stalling radius of the reverse
shock, $R_{\rm rs}$, of $R_{\rm max}$, and of $\rho_w(R>R_{\rm rs})$.
A bow-shock structure will form, with these modified parameters describing its 
forward region. In the other
directions, there will be larger radii and lower densities,
 but only by factors of order unity.   

We will assume that the CSM gas is neutral, both in front and behind
the reverse shock, based of the following
arguments. First, we consider the effect of the reverse shock on the
CSM gas that has passed through it. Since the pressure of this gas
approximately equals the ICM pressure (Eq. \ref{eq:second}), then  
(Eq. \ref{eq:rhow}) implies that the post-reverse-shock CSM
temperature is
\begin{equation}
T_{\rm CSM}\approx 10^4 K v_{w,1.5}^2.
\end{equation}
Thus, the relatively mild 
reverse shock may ionize the gas
at $R>R_{\rm rs}$, but will not heat it excessively. 
If the gas gets ionized, its recombination time will be
\begin{equation}
t_{\rm rec}\approx 1.5\times 10^4 {\rm yr}~v_{w,1.5}^2~ n_{-3}^{-1}~T_1^{-1}, 
\end{equation}
much shorter than the $\sim 10^6$~yr duration of the wind phase. Thus, by
the time the SN explodes, the CSM will have become neutral again.

 Second, the radiation from the ICM also cannot photoionize the CSM.
The photoionization
cross section of neutral hydrogen 
declines with increasing frequency as $\nu^{-3}$
above the ionization threshold, while the optically thin thermal bremsstrahlung
flux from the ICM, per unit frequency, is independent of frequency for
$h\nu\ll kT$. Thus, only photons with energy within
a factor of two above 13.6~eV contribute to ionizations.
For the fiducial ICM parameters, $n_{-3}=1$ and $T_1=1$, 
and assuming a cluster core radius of $\sim 300$~kpc, the 
 bremsstrahlung impinging on the CSM surface has a photon flux
of $\sim 4\times 10^3~{\rm cm}^{-2}~{\rm s}^{-1}$ between 
13.6~eV, and 27.2~eV (e.g., Rybicki \&
Lightman 1979). Multiplying by the ionization cross-section at
13.6~eV,
$6.3\times 10^{-18}~{\rm cm}^2$, the ionization rate is
$\sim 2\times 10^{-14}~{\rm s}^{-1}$ per hydrogen atom.
For the fiducial CSM density of $\sim 10~{\rm cm}^{-3}$ (Eq. \ref{eq:rhow}),
the recombination rate is
$\sim 2\times 10^{-12}~{\rm s}^{-1}$, 100 times higher than the
photoionization rate, and therefore the CSM gas will remain mostly neutral. 
Finally, the ICM gas particles cannot collisionally heat and ionize the CSM 
because magnetic fields
keep their mean free path small, thus suppressing heat
conduction. This situation is actually observed in the case of 
``cold fronts'', clouds of relatively cool gas that
 maintain their lower temperature as they
plunge through an X-ray cluster (e.g., Vikhlinin \& Markevitch 2002).
The magnetic fields also provide surface tension that prevents
shearing of the cold gas clous via the Kelvin-Helmholtz instability.  

\subsection{Ejecta dynamics and line emission}

Once the SN explodes, the fast SN ejecta expands and drives a strong shock
into the CSM. By definition, the ejecta of type Ia explosions have a low
abundance of hydrogen. The H$\alpha$ emission from their remnants
originates from the neutral hydrogen intercepted by the ejecta's forward
shock in the surrounding medium. The underlying physics of H$\alpha$
emission at the young stage (when the shock is called a ``non-radiative''
shock, and the SNR is called ``Balmer dominated'') is well understood
(e.g., Raymond 1991). Of order one tenth of all hydrogen atoms entering the
shock produce an H$\alpha$ photon before they get ionized, and roughly
another tenth of the atoms do it by charge exchange with the hot electrons
behind the shock. The first component generates a narrow line (since the
emission occurs before the atoms get kicked by the shock), and the second
results in a broad H$\alpha$ component with a velocity width of order the
shock speed.

The total H$\alpha$ luminosity can be calculated from the 
radius of the forward shock, $R_{\rm fs}$,
\begin{equation}
L_{H\alpha}= 4\pi R_{\rm fs}^2 {dR_{\rm fs}\over dt}
\left({\rho_{\rm CSM}\over m_p}\right) \times 0.2\times h\nu_{H\alpha},
\end{equation}
where $h\nu_{H\alpha}=3\times 10^{-12}$ erg is the energy of an H$\alpha$
photon. The CSM mass density at radii $R<R_{\rm rs}$ is given by $\rho_{\rm CSM}={\dot
M}/4\pi R^2v_w$, yielding 
\begin{equation}\label{eq:Lal}
L_{H\alpha}={\dot{R_{\rm fs}}\over v_w}{{\dot
M}\over m_p}0.2 h\nu_{H\alpha}.
\end{equation}

The forward shock velocity may be estimated as follows.  Ignoring, at
first, the interaction of the SN ejecta with the CSM, the time dependent
ejecta velocity profile is given, after significant expansion of the
ejecta, by $v=R/t$.  This velocity profile describes ``free expansion"
(expansion into vacuum of a pressureless fluid), in which the velocity of
each fluid element is independent of time. The density profile of such a
flow is self-similar, $\rho(R,t)\propto g(R/t)t^{-3}=g(v)t^{-3}$.
Numerical models of SN-Ia explosions yield $g(v)=\exp(-v/v_{\rm ej})$
(e.g., H\"{o}flich \& Khokhlov 1996; Dwarkadas \& Chevalier 1998).  
The characteristic ejecta speed, $v_{\rm ej}$,
defines the ratio between the ejecta kinetic energy, $E$, and mass, $M$,
\begin{equation}
\frac{E}{M}=\frac{1}{2}\frac{\int dR R^2 \rho v^2}{\int dR R^2 \rho}= 
\frac{\int dv v^4 g(v)}{2\int dv
v^2 g(v)}. 
\end{equation}
For $g(v)=\exp(-v/v_{\rm ej})$ we have
\begin{equation}\label{eq:v_ej}
    v_{\rm ej}=\left(\frac{E}{6M_{\rm ej}}\right)^{1/2}
    =10^{3.5}\left(\frac{E_{51}}{M_{\rm ej,0}}\right)^{1/2}\,{\rm km~s}^{-1},
\end{equation}
where $E=10^{51}E_{51}$~erg and $M_{\rm ej}=10^0M_{\rm ej,0}M_\odot$ (see,
e.g., Truelove \& McKee 1999, for a summary of the values of these
parameters in historical SNRs).  The fraction of the ejecta mass which has
a velocity in excess of $v$ is determined by the algebraic relation
\begin{eqnarray}\label{eq:M_v}
\nonumber
    \frac{M(>v)}{M_{\rm ej}}&=&
    \frac{1}{2}\int_{v/v_{\rm ej}}dx\,x^2e^{-x} \\
    &=&\left[1+\left(1+\frac{1}{2}\frac{v}{v_{\rm ej}}
    \right)\frac{v}{v_{\rm ej}}\right]e^{-v/v_{\rm ej}}.
\end{eqnarray}

The part of the ejecta with velocity $>v$ starts decelerating once it
has propagated to
a radius out to which the CSM mass is comparable to $M(>v)$. As long as 
$R_{\rm fs}<R_{\rm rs}$, the time at which this deceleration occurs is given by
$t=M(>v) v_w/\dot{M}v$. Thus, the velocity of the ``piston" driving the shock into
the surrounding CSM is given, as a function of time, by
\begin{equation}\label{eq:v_t}
    t=\frac{M_{\rm ej}v_w}{\dot{M}v_{\rm ej}}\left[\frac{v_{\rm ej}}{v}+
    \left(1+\frac{1}{2}\frac{v}{v_{\rm ej}}
    \right)\right]e^{-v/v_{\rm ej}}.
\end{equation}
Defining a characteristic time for deceleration,
\begin{equation}\label{eq:t_ej}
    t_{\rm ej}\equiv\frac{M_{\rm ej}v_w}{\dot{M}v_{\rm ej}}=
    1.1\times10^4\frac{M_{\rm ej,0}^{3/2}v_{w,1.5}}{\dot{M}_{-6}E_{51}^{1/2}}\,{\rm yr},
\end{equation}
the solution of equation~(\ref{eq:v_t}) for $t< t_{\rm ej}$ is
\begin{equation}\label{eq:v_t_app}
    v\approx1.1\ln(2t_{\rm ej}/t)v_{\rm ej}.
\end{equation}
Equations~(\ref{eq:v_t}) and (\ref{eq:t_ej}) hold for $R_{\rm fs}<R_{\rm
rs}$. The shock first reaches this radius at $t=t_{\rm rs}$, given by
\begin{equation}\label{eq:trs}
t_{\rm rs}\ln\frac{2t_{\rm ej}}{t_{\rm rs}}\equiv {R_{\rm rs}\over 1.1v_{\rm ej}}= 
58~\left(\frac{{\dot M}_{-6}v_{w,1.5}M_{\rm ej,0}}{n_{-3}T_{*1}E_{51}}\right)^{1/2}~{\rm yr}.
\end{equation}
Since the forward shock is strong, its speed
is simply related to the piston speed by $(dR_{\rm fs}/dt)={\gamma+1\over
2}v={4\over
3}v$, and consequently
\begin{equation}
%\label{eq:CSMLal}
L_{H\alpha}= 
3.0\times 10^{33}~\frac{{\dot M}_{-6}E_{51}^{1/2}}{v_{w,1.5}M_{\rm ej,0}^{1/2}} \ln\frac{2t_{\rm ej}}{t}
~{\rm erg~s^{-1}}
\end{equation}
for $t<t_{\rm rs}$.

For $t>t_{\rm rs}$, $R_{\rm fs}>R_{\rm rs}$, the piston velocity and time are related by
$M(>v)=4\pi(vt)^3\rho_w(R>R_{\rm rs})/3=(4/3)\dot{M}(vt)^3/R_{\rm rs}^2v_w$, i.e. by
\begin{equation}\label{eq:v_t1}
    \left(\frac{t}{t'_{\rm ej}}\right)^3=\left[\left(\frac{v_{\rm ej}}{v}\right)^3+
    \left(1+\frac{1}{2}\frac{v}{v_{\rm ej}}
    \right)\left(\frac{v_{\rm ej}}{v}\right)^2\right]e^{-v/v_{\rm ej}},
\end{equation}
with
\begin{equation}\label{eq:t_ej1}
    t'_{\rm ej}\equiv\left(\frac{3M_{\rm ej}v_wR_{\rm rs}^2}{4\dot{M}v_{\rm ej}^3}\right)^{1/3}=
    360\left(\frac{M_{\rm ej,0}^{5/2}v_{w,1.5}^2}{E_{51}^{3/2}n_{-3}T_{*1}}\right)^{1/3}\,{\rm yr}.
\end{equation}
At $t\le t'_{\rm ej}$ we have $v\approx v_{\rm ej}$. Here, we neglect the
logarithmic corrections to $v$ implied by equation~(\ref{eq:v_t1}), since
for a uniform $\rho_{\rm CSM}$ the time dependence of $L_{H\alpha}$,
$L_{H\alpha}\propto R_{\rm fs}^2v\rho_{\rm CSM}\propto v t^2$, and the
logarithmic evolution of $v$ is not important. We therefore obtain
\begin{equation}
\label{eq:CSMLal}
L_{H\alpha}= 
2.8\times 10^{35}M_{\rm ej,0}^{1/6}E_{51}^{1/2}\left(\frac{n_{-3}T_{*1}}{v_{w,1.5}^2}\right)^{1/3}\left(\frac{t}{t'_{\rm ej}}\right)^\beta
~{\rm erg~s^{-1}}
\end{equation}
where $\beta=2$ for ${R_{\rm rs}/v_{\rm ej}}<t<t'_{\rm ej}$ 
and $\beta=1/5$ for $t'_{\rm
ej}<t$. At $t'_{\rm ej}<t$, equation~(\ref{eq:v_t1}) (which implies
$v\approx v_{\rm ej}(t'_{\rm ej}/t)$) no longer holds, and the shock
approaches the self-similar Sedov-von Neumann-Taylor regime with $R_{\rm
fs}\propto t^{2/5}$.

Equation~(\ref{eq:CSMLal}) holds up to the time where the ejecta traverses
the CSM and enters the ICM. For $v_{\rm ej}t'_{\rm ej}>R_{\rm max}$ the
ejecta suffers little deceleration prior to crossing the CSM, and hence the
crossing time is given by $t_{\rm CSM}\sim R_{\rm max}/ v_{\rm ej}$. For
$v_{\rm ej}t'_{\rm ej}<R_{\rm max}$, the ejecta enters the self-similar
deceleration phase before crossing the CSM and $t_{\rm CSM}\sim (R_{\rm
max}/v_{\rm ej})(R_{\rm max}/v_{\rm ej}t'_{\rm ej})^{3/2}$. Since $v_{\rm
ej}t'_{\rm ej}/R_{\rm max}=(M_{\rm ej}/\dot{M}\tau)^{1/3}$, we find
\begin{eqnarray}
\nonumber
t_{\rm CSM}&\sim& {R_{\rm max}\over v_{\rm ej}}\max\left[1,\left(\frac{\dot{M}\tau}{M_{\rm ej}}\right)^{1/2}\right]
\\ \nonumber &=& 360\left(\frac{{\dot M}_{-6}v_{w,1.5}^2\tau_6}{n_{-3}T_{*1}}\right)^{1/3}\left(\frac{E_{51}}{M_{\rm ej,0}}\right)^{-1/2}
\\ &\times&\max\left[1,\left(\frac{\dot{M}\tau}{M_{\rm ej}}\right)^{1/2}\right]{\rm yr}.
\label{eq:CSM}
\end{eqnarray}

This time may change by a factor of
order unity, depending on the binary speed $v_{\rm star}$, and one could
average over a Maxwellian distribution
 of possible stellar velocities to examine the
statistics of SN events with different durations.  Depending on the
giant's wind speed, the duration of mass loss, and the ejecta velocity, the
time to traverse the CSM may vary in the range $\sim 10^2$--$10^3$ years.

As long as the ejecta is interacting with the CSM, an intergalactic SNR is
thus similar to a galactic SNR at this stage of its development. We can
therefore test the applicability of our simple model by comparing its
predictions to the observed properties of young ($\lesssim 1000$~yr)
type-Ia SNRs in the Milky Way and in nearby galaxies. The remnants of the
historical SNe of 1006 and of 1572 (Tycho) belong to this class.
Observational estimates of sizes and luminosities for Galactic SNRs are
complicated by uncertainties in distance and extinction, but in recent
years this problem is being overcome by the detection of SNRs in nearby
galaxies. In the Large Magellanic Cloud (LMC), the typical H$\alpha$
luminosities of young, Balmer-dominated, SNRs are of order
$10^{34-35}$~erg~s$^{-1}$ (Tuohey et al. 1982; Smith et al. 1991). This is
comparable to our estimate in equation~(\ref{eq:CSMLal}). Note that the
typical pressures, $nT$, of the ICM and the ISM are similar, and hence in
view of the weak, 1/3 power, dependence on pressure, we expect similar
luminosities at this stage from galactic SNRs and intergalactic SNRs
exploding into a CSM.

In older (a few thousand to 20,000 years old) Galactic and nearby SNRs, the
radii are typically 10-15~pc, and H$\alpha$ luminosities are in the range
of $10^{35-37}$~erg~s$^{-1}$ (e.g., Blair \& Long 2004; Williams et
al. 2004). However, at this stage, the shocks, now called ``radiative
shocks'', have slowed down considerably. Cooling is dominated by hydrogen
recombination and by collisionally excited lines of low-ionization metals,
especially in regions where the shock encounters denser clumps with shorter
recombination times.  In fact, optical surveys often distinguish SNRs from
H~II regions based on an emission-line ratio criterion of
[S~II]$\lambda\lambda 6717, 6731$/H$\alpha > 0.4$, but of course, this will
select against young, often Balmer-dominated, SNRs.  It is possible that
intergalactic SNRs can approach this stage of greater Balmer line
luminosity and a metal-line-cooling spectrum in regions of higher pressure
in the ICM (e.g., near the cluster center), or if the pre-explosion wind
included clumps of dense material. More likely, however, the forward shock
will reach $R_{\rm max}$ and enter the ICM when it is still fast and
non-radiative.  Since the ICM is fully ionized, the H$\alpha$ emission is
expected to diminish as soon as the ejecta's forward shock traverses the
CSM. This is in contrast to SNRs in the ISM of galaxies, where the forward
shock continues to intercept neutral atoms even beyond the CSM.

\subsection{Continuum emission}

The collisionless forward shock of the SN remnant is expected to accelerate
electrons to relativistic energies by the Fermi mechanism
(e.g., Blandford \& Eichler 1987).  
In the presence of intracluster magnetic fields, the
accelerated electrons are expected to emit non-thermal radio photons via
the synchrotron process (see Chevalier \& Raymond 1978, and references therein). Let us
first consider the shock driven into the ICM, at the late SNR stage. 
In analogy with SNRs
in galaxies, we assume that a fraction $\xi_e$ of the post-shock thermal
energy is given to the relativistic electrons. For a strong shock, the
accelerated electron number is expected to be distributed with Lorentz
factor $\gamma$ as $dN_e/d\gamma \propto \gamma^{-2}$ up to a maximum
Lorentz factor, $\gamma_{\rm max}$.  The adiabatic compression of the
magnetic field in X-ray clusters is expected to produce a magnetic field
strength of $B\sim 10 B_{-5}~\mu{\rm G}$ in the gas behind the forward
shock front. The accelerated electrons will emit synchrotron radiation at a
frequency,
\begin{equation}
\nu =\gamma^2\left({eB\over 2\pi m_ec}\right)=0.3B_{-5}\gamma_4^2~{\rm GHz},
\end{equation}
where $\gamma_4=(\gamma/10^4)$.  The synchrotron cooling time of the
electrons emitting at a frequency $\sim \nu$ is $t_{\rm syn}\sim
2.5\times 10^7~{\rm yr}~B_{-5}^2\gamma_4^{-1}$.  The synchrotron emission
spans up to $\sim 2\ln \gamma_{\rm max}$ decades in frequency, and the
electrons carry an equal amount of energy per logarithmic Lorentz factor
interval. Assuming that the ejecta does not decelerate significantly
during the preceding 
CSM crossing, which holds for $\dot{M}\tau<M_{\rm ej}$, the synchrotron 
luminosity is therefore given by
\begin{equation}
\nu L_\nu \sim \left({\xi_e
 \over 2\ln\gamma_{\rm max}}\right)\times 
\left({3\times {1\over 6}v_{\rm ej}^2
\times {4\pi\over 3} R^3 
\rho_{\rm ICM}}\right) t_{\rm syn}^{-1}.
\end{equation}
Substituting $\gamma_{\rm max}\sim 10^{10}$ and $\xi_e\sim 0.05$,
based on SNR observations (e.g. Dyer et al. 2001, Ellison et al. 2001; for a discussion see
Keshet et al. 2003), we find that during the passage of the
ejecta through the surrounding ICM, but before the ejecta starts to
decelerate, the radio luminosity is given by
\begin{equation}
\label{eq:ICMradio}
\nu L_\nu\sim 6\times 10^{32}~\left({\nu\over 1~{\rm
GHz}}\right)^{1/2}n_{-3}v_{\rm ej,3.5}^5~t_4^3~{\rm erg~s^{-1}},
\end{equation}
where $t_4=(t/10^4~{\rm yr})$. Equation~(\ref{eq:ICMradio}) holds until the
ejecta begins to decelerate, at
\begin{equation}\label{eq:t_dec}
    t_{\rm dec}\simeq10^4\left(\frac{M_{\rm ej,0}}{n_{-3}}\right)^{1/3}v_{\rm ej,3.5}^{-1}{\rm yr}.
\end{equation}
The luminosities given in equation~(\ref{eq:ICMradio}) are consistent with
those measured in galactic radio SNRs -- $10^{32-34}{\rm erg~s^{-1}}$
(e.g., Payne et al. 2004; Warren \& Hughes 2004) -- for which
$n_{-3}\sim10-1000$, and $t_4\sim 0.1-1$.

The continuum emission (accompanied by a sub-dominant inverse-Compton
component of upscattered microwave background photons), extends to high
frequencies, following equation~(\ref{eq:ICMradio}), up to optical
frequencies, where the cooling time of electrons becomes comparable to the
dynamical time. At higher frequencies the luminosity is frequency
independent, $\nu L_\nu\propto\nu^0$. The peak luminosity is achieved at
the deceleration time, $t=t_{\rm dec}$, when the flux and spectrum are
given by
\begin{eqnarray}
\label{eq:Lmax}
\nonumber \nu L_\nu&\sim& 1.3\times 10^{36}M^{2/3}_{\rm ej, 0.6}v_{\rm
ej,3.5}^3n_{-3}^{1/3} \\ &\times& \left(\frac{\nu/10^{15}{\rm
Hz}}{n_{-3}^{2/3}v_{\rm ej,3.5}^2M_{\rm ej,0.6}^{-2/3}}\right)^{\alpha}~{\rm
erg~s^{-1}},
\end{eqnarray}
where $\alpha=1/2$ for $\nu<10^{15}n_{-3}^{2/3}v_{\rm ej,3.5}^2M_{\rm
  ej, 0.6}^{-2/3}{\rm Hz}$ and $\alpha=0$ otherwise.
This continuum emission can be searched for at optical and X-ray wavelengths.

Looking back now to the
expansion of the ejecta through the CSM, the emission of synchrotron
radiation from the forward collisionless shock driven into the wind depends
on the fraction $\xi_B$ of post shock thermal energy carried by the
magnetic field. Radio supernovae are commonly modelled assuming a
near-equipartition magnetic field, $\xi_B\sim0.1$ (e.g., Weiler et
al. 1998). Under this
assumption, the synchrotron emission during the CSM crossing phase is given
by
\begin{eqnarray}
\label{eq:CSMradio}
\nonumber\nu L_\nu\sim 6&\times& 10^{33}~{\rm erg}{\rm
s}^{-1}~\left({\nu\over 1~{\rm
GHz}}\right)^{1/2}\xi_{B,-1}^{3/4}\left(\frac{n_{-3}T_{*1}}{v_{\rm
w,1.5}^2}\right)^{7/4} \\ & \times & v_{\rm ej,3.5}^{13/2} t_2^3 
\times \cases{ 1, &
$t>t_{\rm rs};$ \cr \left(\frac{R_{\rm rs}/v_{\rm
ej}}{2t}\right)^{7/2}\left[\ln\left(\frac{2t_{\rm ej}}{t}\right)\right]^3,
& $t<t_{\rm rs}$ }~,
\end{eqnarray}
where $\xi_{B}=0.1\xi_{B,-1}$ and $t=10^2t_2$~yr.  The logarithmic factor
cubed is due to the fact that part of the ejecta moves at a velocity $v$
faster than $v_{ej}$. Although this factor formally diverges at small $t$,
$v/v_{ej}$ should not exceed a factor of 10, since the simulations for
various models do not show shells moving at $v>10v_{ej}$
(e.g., H\"{o}flich \& Khokhlov 1996; Dwarkadas \& Chevalier 1998).  
Since $v/v_{ej}\gtrsim\ln(2t_{\rm ej}/2)\sim10$
is obtained for $t\lesssim1$~yr, equation (\ref{eq:CSMradio}) is not
applicable at earlier times.  For the fiducial parameters, the 1~GHz
luminosity $\sim 1$ year after the explosion is $\nu L_{\nu}\sim 1.4\times
10^{35}$~erg~s$^{-1}$.  It should be kept in mind that, at this stage, the
luminosity depends sensitively on the velocity distribution in the ejecta
-- if the maximum velocity is, say, $5 v_{ej}$, then the luminosity will be
be 10 times smaller.

Observationally, prompt or very early (few years) radio emission from
type-Ia SNe has not been detected. Sramek \& Weiler (1990; see Boffi \&
Branch 1995) have presented upper limits on the radio fluxes for several
events at distances of $\sim 20$~Mpc, which translate to luminosity limits
of $\nu L_{\nu}\lesssim 10^{35}$~erg~s$^{-1}$.  Three events were observed
$\sim 2$~months after the explosion, but one event, SN1981B, was observed
near optical maximum. A $3\sigma$ upper limit of 1mJy at 5~GHz was also
obtained a week before optical maximum by Eck et al. (1995) for SN1986G in
the nearby (4.2~Mpc; Tonry et al. 2001) galaxy NGC~5128 (Cen A),
corresponding to $\nu L_{\nu}< 1.1\times 10^{35}$~erg~s$^{-1}$.  However
1986G was a peculiar and underluminous SN-Ia.  These upper limits are
comparable to the radio luminosities that we predict at an age $\sim
1$~yr. Given the freedom in input parameters to equation
(\ref{eq:CSMradio}), the dependence of its range of applicability on the
speed of the fastest ejecta, and the small number of observed cases, this
does not yet pose a serious discrepancy. However, tighter upper limits on
the radio flux from additional SNe-Ia would constitute independent evidence
for the absence of a CSM. Similar conclusions were reached by Boffi \&
Branch (1995).

\section{Have Optical Intergalactic SNRs Been Detected?}

In the course of 
 narrow-band imaging of the Virgo cluster in search of intracluster
planetary nebulae, Gerhard et al. (2002) have recently discovered a
detached, spatially unresolved, emission-line object, $\sim 17$~kpc in
projection from the spiral galaxy NGC~4388.  Optical spectroscopy of the
object revealed line ratios, including [S~II]$\lambda\lambda 6717,
6731$/H$\alpha =0.1$, which are characteristic of H~II regions. Based on
this, Gerhard et al. (2002) classified this object as an intergalactic H~II
region, in which the line emission is powered by one or two O stars. The
massive stars are presumably members of a young cluster that was formed
{\it in situ} in the ICM as a result of a recent collision between
galaxies. The H$\alpha$ equivalent width constrains the age of the star
cluster to $\sim 3$~Myr, too short for the stars to have formed in NGC~4388
and to have traversed the large distance, although the similar radial
velocities of the galaxy and the emission-line region do suggest that the
two are associated.  The physical mechanism that could have led to
intergalactic star formation is unclear, but perhaps the galaxy collision
caused stripping and compression of gas along tidal tails, which then
fragmented into stars when the gas was already unbound from its original
galaxy.  Gerhard et al. (2002) found a total of 17 candidate objects of
this type in their narrow-band data, with H$\alpha$+[N~II] luminosities of
order $10^{37}$~erg~s$^{-1}$ (for an assumed Virgo distance of 17~Mpc;
since the exact distances are unknown, the luminosities could be lower
or higher by an order of magnitude).

Several, possibly related, objects have also been found by Ryan-Weber et
al. (2004)
in a narrow-band imaging survey around nearby galaxies.
Again, the emission-line objects are
 at large projected distances from their associated
galaxies, but at similar velocities.
 Two of these candidates have been spectroscopically
confirmed to be low-redshift objects based on the detection of both
H$\alpha$ and [OIII]$\lambda 5007$ (the signal-to-noise ratio
 was too low for detection of
additional lines). One object is 33~kpc from
the S0 galaxy NGC~1533 in the Doradus Group (distance 21~Mpc), 
and another is 19~kpc from 
the galaxies NGC~833 and NGC~835 in the compact group HCG~16 (distance
53~Mpc).
The H$\alpha$ luminosities, and the deduced ages and numbers of ionizing
stars are of the same order of magnitude as found by Gerhard et
al. (2002) in Virgo. 
     
While the intergalactic star-formation option is a possibility, we 
point out that, within current observational constraints, the emission
line objects could also be the remnants of the intracluster type-Ia
SN population discovered by Gal-Yam et al. (2003), the observational
signatures of which we have estimated above.  
The angular scale occupied by a remnant at the distance of the Virgo
cluster, $d=10d_{1}~$Mpc, is
\begin{equation}
\theta \sim {R_{\rm max}\over d}= 0\farcs02
\left({R_{\rm 18.5} \over d_1}\right),
\end{equation}
i.e., unresolved by ground-based optical telescopes, but resolvable with
the Hubble Space Telescope (HST) if $R_{\rm 18.5}\gtrsim 3$.  The H$\alpha$
luminosities of the putative intergalactic H~II regions are similar to
those of the more luminous SNRs in nearby galaxies (e.g., Blair \& Long
2004).
%Detection of objects of such low luminosity at Virgo distances and
%beyond is challenging, and it is therefore natural that the upper end of
%the luminosity distribution would have been detected.  
Nearby SNRs of this luminosity are at least several thousand years old, and
are in their slow, radiative-shock, phase, with strong low-ionization metal
lines, in addition to Balmer lines.  We estimated above that, under typical
conditions, intergalactic SNRs will not reach this phase. Furthermore, the
object found by Gerhard et al. (2003) has detectable but relatively weak,
metal lines.  However, considering the distance uncertainties (which lead
to an order of magnitude uncertainty in the luminosity), stretching the
parameters in equation (\ref{eq:CSMLal}), and allowing for line emission
from dense clumps in the giant's wind, it cannot be excluded that this
object is an intergalactic SNR.  In the case of the objects studied by
Ryan-Weber et al. (2004), the spectra have too low a signal-to-noise ratio
to detect the metal lines, let alone perform the usual diagnostic tests
distinguishing H~II regions from SNRs.

In terms of numbers, Gerhard et al. (2003) estimate that there are $\sim
10^3$ similar emission-line objects in Virgo, but of order one-half, or
even more, of these sources are likely background objects at high redshift
(see Ryan-Weber et al. 2004). The measured type-Ia SN rate, per unit
$B$-band stellar luminosity, in clusters and in elliptical galaxies (see
summary in Gal-Yam et al. 2002) is $R=0.2\pm 0.1 h^2_{70}$~SNu, where
1SNu$=$SN century$^{-1} (10^{10}L_{B,\odot})^{-1}$, and $h_{70}$ is the
Hubble parameter in units of 70~km~s$^{-1}$~Mpc$^{-1}$. The expected number
of intergalactic SNRs in Virgo is therefore
\begin{equation}
\label{eq:numSNRs}
N_{\rm SNR}\sim R f L t\sim 150 R_{0.2} f_{0.2} L_{12.5} t_{3},   
\end{equation}
where $R_{0.2}=R/(0.2~{\rm SNu})$, $f=0.2 f_{0.2}$ 
is the intergalactic stellar fraction, $L=3\times 10^{12}L_{B,\odot} L_{12.5}$
is the total stellar $B$-band luminosity of Virgo (Sandage et
al. 1985; Trentham \& Hodgkin 2002), and 
$t=10^3~{\rm yr}~t_{3}$
is the time during which a SNR emits H$\alpha$ via non-radiative
shocks that encounter the CSM. The number of emission-line objects is 
therefore also consistent, to an order of magnitude,
 with the SNR option, if the outer CSM radius
is large enough to keep the SNR bright for about one thousand years.    

Thus, if the intergalactic emission-line objects are SNRs,
high-angular-resolution imaging should reveal resolved, shell-like
morphologies with diameters of order 3~pc.
If these objects are the remnants
 of the intergalactic type-Ia SN population,
this would establish the fact that SNe-Ia have a CSM at the time of
explosion, with far-reaching implications for progenitor models. 

We note that, based on the rate parameters above, there should be an
intergalactic SN-Ia in Virgo about once per decade.  However, such SNe may
be missed because surveys for nearby SNe (e.g., Li et al. 2003) monitor
individual galaxies, rather than the entire cluster.  A candidate
intergalactic type-Ia SN that went off in Virgo is SN1980I (Smith et
al. 1981), which occurred in between three elliptical/S0 galaxies, but was
separated in projection by $\sim 50$~kpc from each. Twenty-five years after
the explosion, the SNR should be about halfway to its CSM reverse-shock
crossing stage, with an H$\alpha$ luminosity of $\sim
10^{34}$~erg~s$^{-1}$, i.e., an H$\alpha$ flux of $\sim 3\times
10^{-19}$~erg~s$^{-1}$~cm$^{-2}$. Detection of such a weak line will
be possible with the next generation of large ($\sim 30$~m) telescopes.

\section{Intergalactic SNRs in the Radio} 

At the distance of the Virgo cluster of $\sim 10~{\rm Mpc}$, the peak radio
luminosities of intergalactic SNRs, lasting of order $10^4 t_4$~yr,
translate to fluxes of $\sim 0.1$mJy at $\nu\sim 1$GHz. Considering the
intergalactic SN rates above (Eq. \ref{eq:numSNRs}), there should be of
order $\sim 10^3 t_4$ radio SNRs visible above this flux limit in Virgo, or
about 10 SNRs per square degree. This is much smaller than the surface
density of the background radio source population. 
A deep VLA survey of the Hubble Deep Field
by Richards (2000) showed that the mean density of background radio sources
brighter than 0.1~mJy is about 3000 per square degree. He found that about
one half of the background sources were spatially extended, above the $\sim
2''$ resolution limit.  Richards et al. (1998) found that the majority of
the radio sources can be associated with luminous galaxies at redshifts
$z\sim 0.1-1$, with mean $R$-band optical AB magnitudes of $\sim 22$, and
generally brighter than 24.5~mag.  Star-forming disk galaxies and AGNs
(generally in early-type galaxies) both contribute to the source counts at
these fluxes, although the relative contribution of each class is unclear.

Intergalactic SNRs in Virgo would have an angular extent of order
$1^{\prime\prime}$, and would not be associated with any background
galaxy. Thus, by surveying one square degree in Virgo, to $0.1$~mJy in the
radio and to 24.5~mag in the optical, one could exclude most of the $\sim
3000$ background radio sources, based on their large sizes, their
association with distant galaxies, or both.  Using follow-up radio
observations, with higher angular resolution, of the remaining candidates,
one could find of order 10 radio SNRs of intracluster SNe, based on their
characteristic morphologies and sizes.  A detection of this population, and
measurements of its properties (SNR sizes, luminosities, spectra), would
provide constraints on the intergalactic SN rate and on SN-Ia progenitors.

Finally, from equation (\ref{eq:CSMradio}), the 1~GHz continuum flux from
the remnant of SN1980I should be $\sim 0.25$~mJy, which is detectable,
although for some parameters the flux could be lower by an order of
magnitude. With a radius of about 1~milliarcsecond, the remnant should be
unresolved.

\section{Conclusions}

We have estimated the properties of the remnants of type-Ia SNe that
explode in the ICM. We have shown that, if such SNe explode into a CSM,
presumably produced by the wind from the donor giant star from which the
white dwarf accreted mass, then intergalactic SNRs are no different from
galactic SNRs, for an age of up to a thousand years. This conclusion
is mostly unaffected by the large speed with which the SNR may travel
through the ICM. Galactic and intergalactic supernovae are distinct after
they have expanded beyond the CSM radius. Since CSM and galaxy-ISM
densities are comparable, galactic SNRs remain luminous emission-line
sources in the ISM stage. Indeed, for any particular galactic SNR, it is
difficult to determine observationally whether the ejecta is interacting
with the CSM or the ISM, or even whether there ever was a
CSM. 

Intergalactic SNRs, in contrast, are optically luminous only during the CSM
stage, if there is one. Once the ejecta reaches the ICM, the SNR quickly
fades in the optical band. As a result, intergalactic SNRs can provide a
unique test for the existence of a CSM around type-Ia SNe. If intracluster
SNRs are found in the optical band, the corollary would be that type-Ia SNe
have a CSM, supporting the progenitor model of accretion from a giant star
companion. Alternatively, a non-detection of these SNRs would imply the
absence of a CSM. This would point to accretion from a main-sequence or
sub-giant companion, or to the ``double-degenerate'' progenitor model,
involving the merger of a white-dwarf pair following the loss of orbital
energy to gravitational radiation.

We have also shown that irrespective of the existence of a CSM, old
intracluster SNRs are detectable in the radio band. Although identifying
them among the more numerous background radio populations is challenging,
one could take advantage of the low optical luminosity of SNRs in the ICM
stage. In contrast with most of the background radio sources, which are
associated with galaxies, the old intergalactic radio SNRs will have no
associated optical source. Thus, a radio search could establish the
existence of the old intracluster SNR population, and an optical search
would provide valuable information on type-Ia SN physics.

Finally, we have speculated that several examples of compact intergalactic
emission-line objects, recently discovered in Virgo (Gerhard et al. 2002)
and in two galaxy groups (Ryan-Weber et al. 2004), could be intergalactic
SNRs, rather than H~II regions, as postulated before. We have argued that
the luminosities and numbers of the newly discovered objects, although
somewhat on the high side, are still roughly consistent with our estimates
for SNRs. Consistency is possible given the simplifications inherent in our
model, the allowed range of input parameters, and the uncertainties in the
parameters describing the observed objects.  High resolution imaging with
HST could potentially discriminate among the two options, based on the
different morphologies of SNRs and H~II regions. Deep pointed observations
of the site of SN1980I, a possible intergalactic SN-Ia in Virgo, could test
our predictions for the early-stage development of intergalactic SNRs.

\acknowledgements 

We thank John Raymond and Jacco Vink for enlightening discussions on
radiative processes in supernova remnants.  This work was supported in part
by NASA grant NAG 5-13292, and by NSF grants AST-0071019, AST-0204514 (for
A.L.). A. L.  is grateful for the kind hospitality of the Einstein Minerva
Center at the Weizmann Institute of Science, where this work started.

\end{document}